\documentclass[12pt,preprint]{aastex}

\shorttitle{MEASUREMENTS OF THE MEAN DIFFUSE GALACTIC LIGHT SPECTRUM IN THE 0.95~$\mu$m TO 1.65$\mu$m BAND FROM CIBER}
\shortauthors{T. Arai}

\begin{document}


\title{MEASUREMENTS OF THE MEAN DIFFUSE GALACTIC LIGHT SPECTRUM IN THE 0.95~$\mu$m TO 1.65$\mu$m BAND FROM CIBER}


\author{ T. Arai\altaffilmark{1} , S. Matsuura\altaffilmark{1} , J. Bock \altaffilmark{2,3}, A. Cooray\altaffilmark{4} , M. G. Kim\altaffilmark{5}, A. Lanz\altaffilmark{2}, \\
D. H. Lee\altaffilmark{6}, H. M. Lee\altaffilmark{5}, J. Smidt\altaffilmark{11, 4}, T. Matsumoto\altaffilmark{1,7}, T. Nakagawa\altaffilmark{1}, Y. Onishi\altaffilmark{1}, \\
P. Korngut\altaffilmark{3,2}, M.Shirahata\altaffilmark{8},
K. Tsumura\altaffilmark{9}, and M. Zemcov \altaffilmark{2,3}}


\altaffiltext{1}{Department of Space Astronomy and Astrophysics, Institute of Space and Astronoutical Science (ISAS), Japan Aerospace Exploration Agency (JAXA), Sagamihara, Kanagawa 252-5210, Japan}
\altaffiltext{2}{Department of Astronomy, California Institute of Technology, Pasadena, CA 91125, USA}
\altaffiltext{3}{Jet Propulsion Laboratory (JPL), National Aeronautics and Space Administration (NASA), Pasadena, CA 91109, USA}
\altaffiltext{4}{Center for Cosmology, University of California, Irvine, Irvine, CA 92697, USA}
\altaffiltext{5}{Department of Physics and Astronomy, Seoul National University, Seoul 151-742, Korea}
\altaffiltext{6}{Korea Astronomy and Space Science Institute (KASI), Daejeon 305-348, Korea}
\altaffiltext{7}{Institute of Astronomy and Astrophysics, Academia Sinica, National Taiwan University, Taipei 10617, Taiwan R. O. C.}
\altaffiltext{8}{National Institutes of Natural Science, National Astronomical Observatory of Japan (NAOJ), Tokyo 181-8588, Japan}
\altaffiltext{9}{Frontier Research Institute for Interdisciplinary Science, Tohoku University, Sendai 980-8578, Japan}
\altaffiltext{10}{Theoretical Division, Los Alamos National Laboratory, Los Alamos, NM 87545, USA}



\begin{abstract}
We report measurements of the Diffuse Galactic Light (DGL) spectrum
in the near-infrared, spanning the wavelength range 0.95-1.65~$\mu$m by the Cosmic Infrared Background ExpeRiment (CIBER). 
Using the low-resolution spectrometer (LRS) calibrated for absolute spectro-photometry, we acquired
long-slit spectral images of the total diffuse sky brightness towards four
high-latitude fields spread over four sounding rocket flights.
To separate the DGL spectrum from the total sky brightness, 
we correlated the spectral images with a 100~$\mu$m intensity map, 
which traces the dust column
density in optically thin regions.
The measured DGL spectrum shows no resolved features and is consistent with other DGL measurements
in the optical and at near-infrared wavelengths longer than 1.8~$\mu$m.
Our result implies that the continuum is consistently reproduced by models of scattered starlight
in the Rayleigh scattering regime with a few large grains. 
\end{abstract}





\section{Introduction}

Diffuse Galactic Light (DGL) arises from stellar radiation scattered by dust in the interstellar medium.
The DGL spectrum includes information on the optical properties of interstellar dust, such as the grain size distribution and composition,
as well as the interstellar radiation field (ISRF).

Historically, 
DGL was originally detected at optical wavelengths and interpreted as starlight scattered by interstellar dust
\citep{1937ApJ....85..213E, 1941ApJ....93...70H, 1969Phy....41..151V, 1960ZA.....50..121E, 1966Natur.209..388W, 1979A&A....78..253M}. 
The interstellar dust has been studied through its emission properties in the far-infrared \citep{1985A&A...145L...7D, 1987A&A...184..269L, 1995A&A...301..873S}, and was mapped by  IRAS \citep{1984ApJ...278L..19L}.

In the optically thin limit, DGL scales with the light absorbed and re-radiated in the far-infrared, with a proportionality that depends on the properties of the dust grains.
This linear correlation between DGL and far-infrared intensity enables us to separate the DGL component from other diffuse emission, such as Zodiacal Light (ZL), sun light scattered by interplanetary dust, Integrated Star Light (ISL) from undetected stars, and the Extra-galactic Background Light (EBL).

A wavelengths shorter than 0.8~$\mu$m, DGL has a significantly shallower spectral slope than the ISRF due to the wavelength-dependent scattering cross-section of the grains \citep{2012ApJ...744..129B}.
This suggests that the DGL spectrum depends on the size distribution of the grains.
The scattering cross-section can be approximated by Rayleigh scattering theory, 
where the size of the dust grains is  smaller than the wavelength. 
Near-infrared measurements may be more sensitive to the size distribution of dust grains than measurements at optical wavelengths.

Although interstellar dust scattering has been studied in several measurements, 
the size distribution is still under discussion. \citet{1995ApJ...444..293K} insisted that the size distribution 
requires grain sizes ranging from 0.003~$\mu$m to 3~$\mu$m with a peak at 0.2~$\mu$m based on the polarization of star light assuming spheroidal dust particles. 
\citet{2001ApJ...548..296W}, hereafter WD01, compared their own model with the observed extinction of starlight, 
and claimed that interstellar dust includes large grains with radii $a$~$>$~0.2~$\mu$m with a half-mass radius a$_{0.5}$~=~0.12~$\mu$m
where \citet{2011piim.book.....D} defines $a_{0.5}$ as 50$\%$ of the mass in grains with a~$>$~$a_{0.5}$.
A model of \citet{2004ApJS..152..211Z}, hereafter ZDA04, consists of a small population of large grains with radii $a$~$>$~0.2~$\mu$m and many small grains, giving a half-mass radius $a_{0.5}$~=~0.06~$\mu$m. 

Measurement of the DGL spectrum in the near-infrared helps determine the size distribution of interstellar dust. 
However, DGL has not been measured from the ground at these wavelengths because DGL is low-surface brightness and much fainter than atmospheric airglow emission that contaminates large spatial scales.
Even at optical wavelengths, ground-based DGL measurements are problematic, 
and suffer from systematic error due to airglow emission.
Thus we measure the near-infrared DGL from space using the sounding rocket-borne Cosmic Infrared Background ExpeRiment (CIBER) \citep{2006NewAR..50..215B}.
We correlate measured brightness in the near-infrared with diffuse thermal dust emission in the far-infrared to extract the DGL measurement.
Our results are the first DGL measurements at these wavelengths.


\section{Cosmic Infrared Background Experiment}

\subsection{Low Resolution Spectrometer}

\begin{deluxetable}{ccrrrrrrrrcrl}
\tabletypesize{\scriptsize}
\rotate
\tablecaption{Our observed fields of the second flight and the fourth flight. 
RA and DEC indicate equatorial longitude and latitude, respectively, while l and b indicate galactic longitude and latitude, respectively.}
\tablewidth{0pt}
\tablehead{\colhead{Field Name} & \colhead{Exposure time [sec]} & \colhead{Altitude [km]} & \colhead{(RA, DEC) [degree]}  & \colhead{(l, b) [degree]} \\}
\startdata
Second flight & & & & \\
\tableline
SWIRE ELAIS-N1& 83 &139-172&(243.069, 55.283)& (84.89, 44.62)  \\
North Ecliptic Pole (NEP)  & 67 &199-220&(270.871, 66.004 & (96.13, 29.81)  \\
Elat10 ($\beta$ = 10 degree) & 9 &245-310&(234.337, -8.466)& (356.88, 46.08) \\
Elat30 ($\beta$ = 30 degree) & 18 &314-295&(223.058, 20.658) & (23.52, 63.31) \\
BOOTES-A & 63 &288-232&(218.806, 35.120)&  (58.76, 66.79)  \\
\tableline
Fourth flight & & & & \\
\tableline
DGL field & 65 & 272-401 &(251.97, 68.85)&  (100.37, 36.17) \\
North Ecliptic Pole (NEP)  & 60 & 425-505 &(270.82, 66.24)  & (96.02, 29.48) \\
Lockman Hole & 50 & 520-558 &(161.23, 58.58)  & (149.07,  51.65) \\
Elat10 ($\beta$ = 10 degree) & 50 &566-557 &(190.49, 8.02) & (295.80, 70.77)  \\
Elat30 ($\beta$ = 30 degree) & 50 &577-562&(193.05, 27.96) & (111.34, 89.15)\\
BOOTES-B & 55 &555-509 &(217.23, 33.18) & (54.90, 68.13) \\
SWIRE ELAIS-N1& 55 &395-275&(242.84, 54.77) & (84.56, 44.64) \\
\enddata
\label{tbl:field24}
\tablecomments{The coordinate systems are based on J2000.}
\end{deluxetable}

\begin{figure}[htbp]
\epsscale{1.0}
\plotone{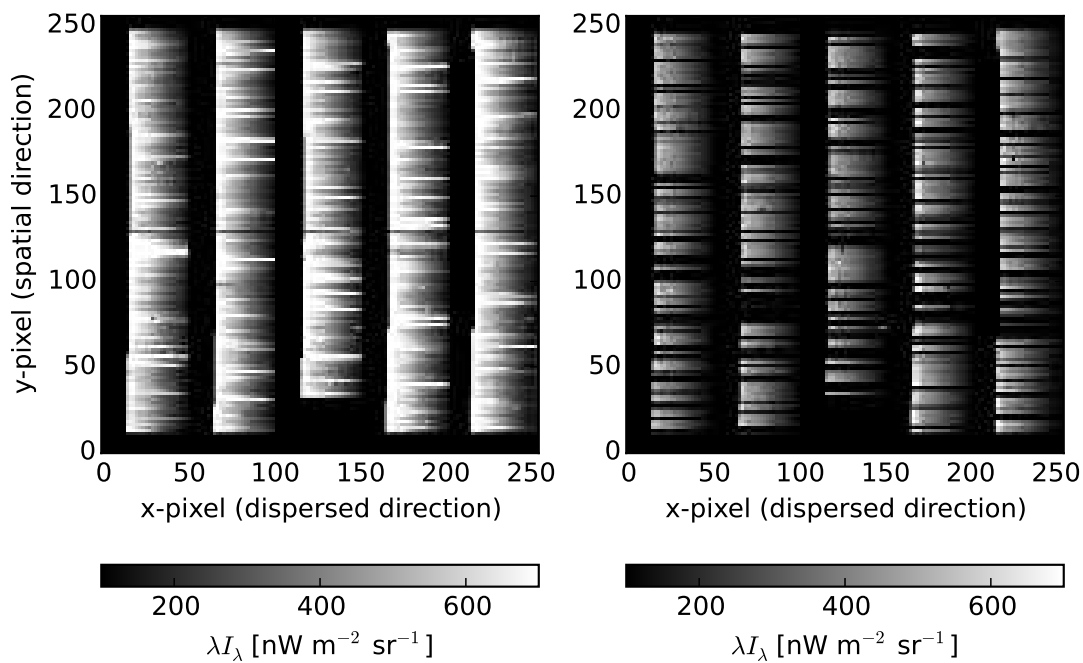}
\caption{Example of an LRS spectral image towards the north ecliptic pole. 
The left panel shows the processed flight image used to derive the DGL spectrum. 
The five vertical sections correspond to the locations of the spectrometer's slits dispersed in the x direction. 
The y direction has spatial information.
Bright stars detected as point sources appear as discrete stripes. 
The top, bottom and left edge of the image are masked, to monitor any short-term changes in dark current. 
The right panel shows the image after subtracting dark current and masking bright stars to isolate the diffuse emission.}
\label{fig:img}
\end{figure}

CIBER is designed to study diffuse near-infrared emission from above the Earth's atmosphere \citep{2013ApJS..207...31Z}.
CIBER has three payload instruments \citep{2013ApJS..207...32B, 2013ApJS..207...34K},
including a Low Resolution Spectrometer (LRS) designed to measure the spectrum of diffuse light in 0.8$\leq \lambda \leq$ 1.8~$\mu$m \citep{2013ApJS..207...33T}.
The LRS consists of an optical collimator which brings an image of the sky to focus on a mask containing 5 slits, 
each spanning a field of view (FOV) of 5$^\circ \times$2.7$'$. 
The light passing through the slits is then dispersed by a prism and brought to focus again on a 256$\times$256 pixel HgCdTe detector array. 

The LRS covers the wavelength range 0.8-1.8~$\mu$m with a resolving power R $= \ \Delta \lambda / \lambda \ =$ 15-30.
The FOV provides a large etendue to measure the diffuse light with high signal to noise ratio (S/N). 
To measure the dark current of the detector, a cold shutter cooled to 77K is mounted just before the detector, and is closed before and during the flight.
In addition, the slit mask provides a masked region where light does not fall on the detector, to monitor any short-term changes in dark current.
The 3$\sigma$ sensitivity of the LRS with 200 spatial pixels in a 50 sec integration is 2.5~nW~m$^{-2}$~sr$^{-1}$ at 1.25~$\mu$m, which enables accurate measurement of the DGL brightness.
Details of the LRS design are described in \citet{2013ApJS..207...33T}.

\subsection{Observations}
CIBER was flown four times, 
in February 2009, July 2010, March 2012 and June 2013.
The payload was successfully recovered and refurbished after the first three flights.
For the fourth flight, a larger launch vehicle was implemented which resulted in a much higher altitude apogee, 
but this configuration did not allow the payload to be recovered.
The observed fields are listed in Table \ref{tbl:field24} with exposure time and altitude.
The first three flights used a two-stage rocket launched from White Sands Missile Range in New Mexico, USA. 
The apogee on these flights was typically 330km, providing a total exposure time of $\sim$~240~seconds.
In the fourth flight, the rocket was launched from Wallops Flight Facility in Virginia, USA, using a four-stage rocket.
 The payload reached 550 km with a total exposure time of 335 seconds.
For this study we use only data from the second and fourth flights, as the first flight data were contaminated by excess stray thermal radiation from the rocket skin, and in the third flight the LRS was operated as a polarimeter.

The raw data, which are non-destructively sampled by the integrating detectors, were telemetered to the ground from the rocket during the flight.
The celestial attitude control system achieved a pointing stability of $<$~8$''$.

\subsection{Field Selection}

\begin{figure}[htbp]
\epsscale{0.6}
\plotone{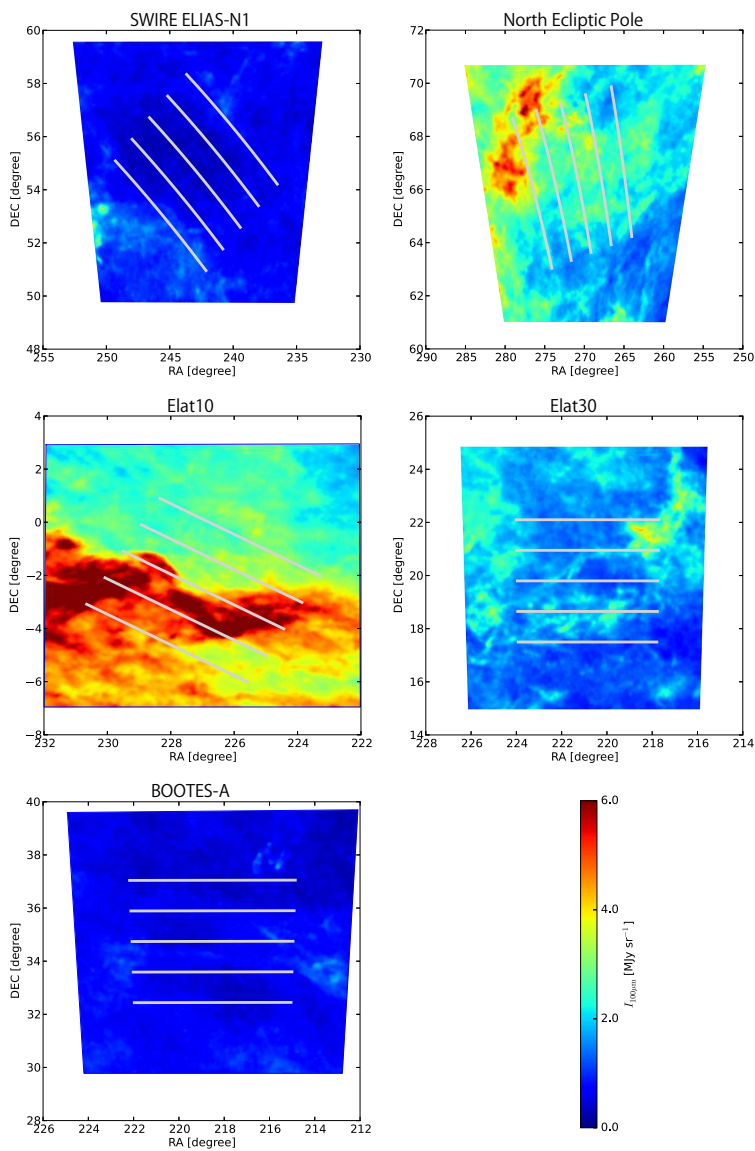}
\caption{The color maps indicate the 100~$\mu$m intensity of the far-infrared cirrus emission \citep{1998ApJ...500..525S} for the second flight. 
The gray lines present the field of view of the LRS.
The equatorial coordinate systems are based on J2000.}
\label{fig:100um_map_2nd}
\end{figure}

\begin{figure}[htbp]
\epsscale{0.6}
\plotone{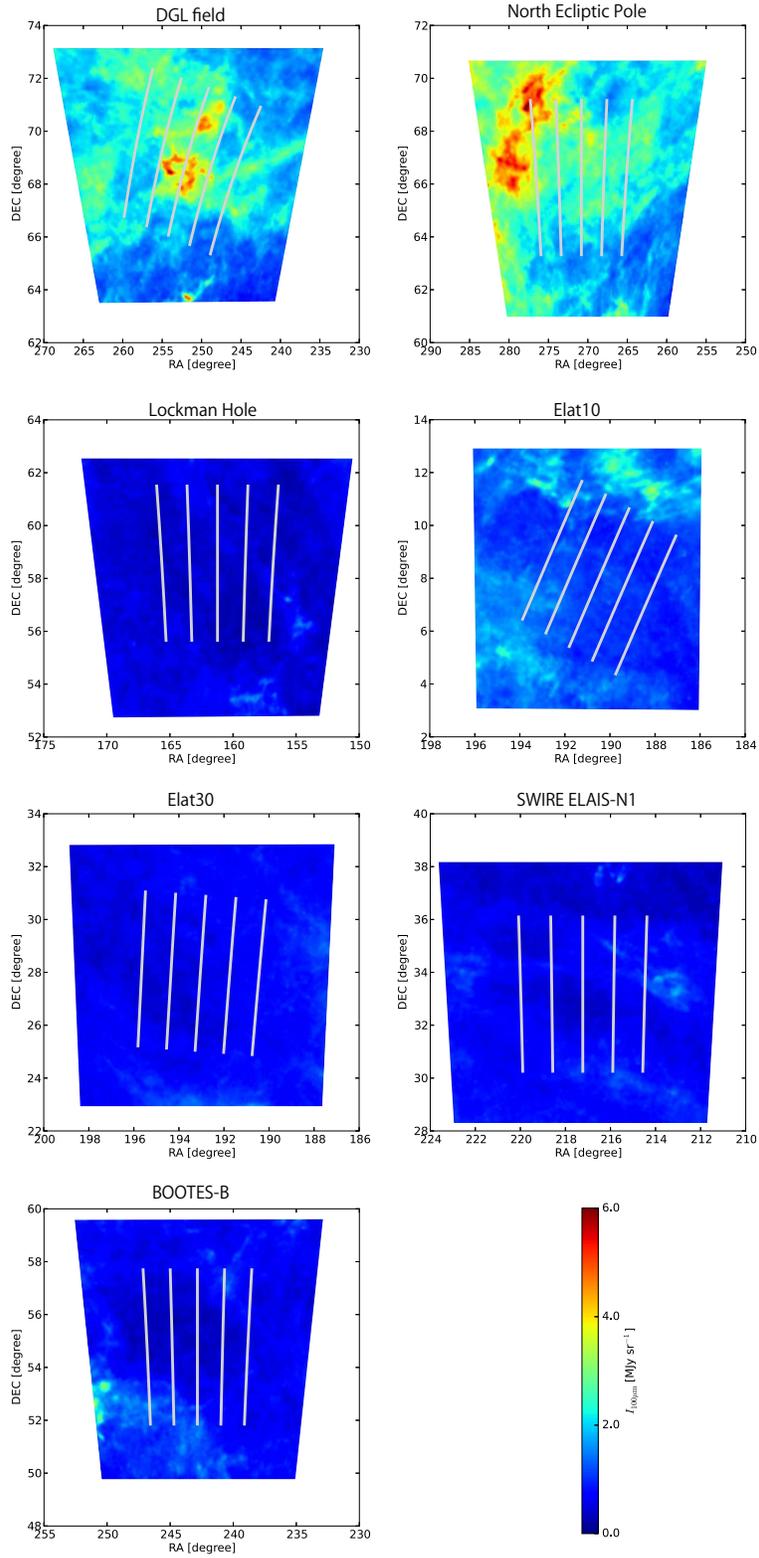}
\caption{As Figure \ref{fig:100um_map_2nd} for the fourth flight.}
\label{fig:100um_map_4th}
\end{figure}

For the DGL analysis, fields with large contrast in $I_{100 \mu {\rm m}}$ emission across an LRS FOV are selected.
These include the NEP field, observed in both the second and fourth flights, 
a field referred to as DGL, specifically targeted for its large expected dynamic range in DGL, 
and the Elat10 field, observed in the second flight. 
Figures \ref{fig:100um_map_2nd} and \ref{fig:100um_map_4th} show the 100~$\mu$m intensity map \citep{1998ApJ...500..525S} of all the fields measured in the second and fourth flights.
The  NEP and DGL fields are high ecliptic latitude fields (see fourth column of Table \ref{tbl:field24}) and have low ZL brightness.
We also select low ecliptic latitude fields with bright ZL, Elat10 and Elat30, in order to make a template spectrum for ZL separation in the observed data.

In all the flights, there is emission from terrestrial atmospheric airglow, 
exospheric atmospheric airglow, and dissociated water vapor that outgasses from the payload early in the flight. 
Although airglow emission dominates the observed spectrum from 1.5~$\mu$m to 1.7~$\mu$m in the first field of every flight,
airglow emission decays exponentially with time and altitude \citep{2010ApJ...719..394T} and is negligible for the remaining fields.
Since the airglow emission is assumed to have no spatial fluctuation in the FOV of the LRS,
we do not subtract it in our DGL measurement.
We discuss the systematic error due to airglow emission in Section \ref{seq:sys}.

\section{Instrument Calibration}

\subsection{Wavelength Calibration}

\begin{figure}[htbp]
\epsscale{0.5}
\plotone{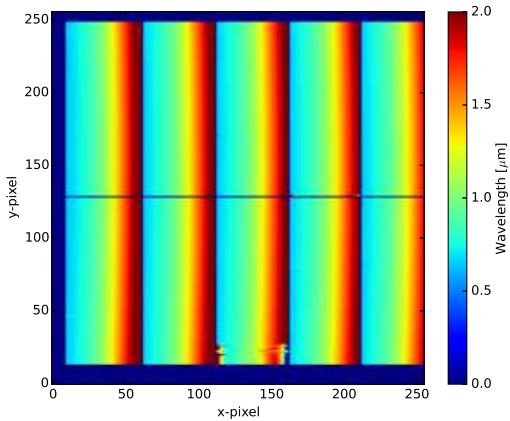}
\caption{Relation between wavelength of the incident light and the position of the detector pixels for the second flight.}
\label{fig:wavemap}
\end{figure}

In order to measure the electromagnetic spectrum of diffuse light,
it is essential to calibrate the wavelength response,
which encodes the mapping of each wavelength of incident light to a position on the detector array.
This is determined through a series of laboratory measurements.
Two different light sources are used for spectral calibration consisting of the SIRCUS (the Spectral Irradiance and Radiance Calibrations using Uniform Sources facility, \citet{2006ApOpt..45.8218B}) laser facility and a standard quartz-tungsten-halogen lamp coupled to a monochrometer.
The wavelength of the SIRCUS laser is determined with an external wavemeter calibrated by National Institute of Standards and Technology (NIST).
The wavelength of the monochrometer is calibrated using a He-Ne laser and the spectral lines of a Ne lamp.
In both cases, monochromatic light was coupled to an integrating sphere through a fiber which illuminated the LRS aperture.
Following each exposure with a monochromic light source, 
we fit the detected signal with a Gaussian function.
The center of this Gaussian, in combination with the externally determined wavelength of incident light, is used to generate the wavelength map shown in Figure \ref{fig:wavemap}.
The wavelength accuracy of the input light is better than 1nm, sufficiently small compared to the $>$~10~nm wavelength resolution of the LRS.

\subsection{Surface Brightness Calibration}\label{sec:sb_cal}

\begin{figure}[htbp]
\epsscale{0.5}
\plotone{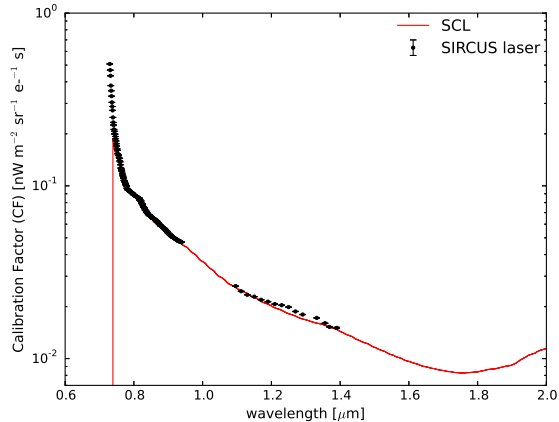}
\caption{Comparison of the calibration factor obtained with the SIRCUS laser and the super-continuum laser.
The red line indicates the calibration factor of super continuum laser. The black dots give the calibration factor of the SIRCUS laser.}
\label{fig:cf}
\end{figure}

Absolute spectro-photometric calibration is done through laboratory measurements, using the techniques described in \citet{2013ApJS..207...31Z}. 
Two different light sources are utilized in this calibration measurement: a super-continuum laser (SCL) for broad-band measurement, 
and the SIRCUS laser facility. 
These light sources are coupled to a 4" diameter aperture integrating sphere \footnote{Manufactured by Gigahertz-Optik, Inc.} whose port is viewed by the LRS. 
The absolute brightness of the integrating spheres viewed by the LRS is determined using absolutely calibrated radiometers and a monitor detector. 
During the measurements, the light sources are shuttered and the measured ambient signal is subtracted from the data.

The dynamic range mismatch between the radiometer and the LRS requires a two-step bootstrapping approach.
To avoid non-linearity effects in the LRS detector array introduced at high-photocurrent levels,
the brightness of the integrating sphere must be attenuated.
The absolutely calibrated radiometers measure the absolute brightness of the integrating sphere with light levels 10$^{4}$~$\sim$~10$^{6}$ brighter than for the LRS measurement. 
The monitor detector measures the intensity of a secondary smaller injection sphere with sufficient S/N in both cases to measure the coupling factor between the two measurement points.
Therefore, when the intensity is reduced to levels accessible to the LRS, 
the absolute brightness of the main sphere is inferred using the monitor signal and the previously determined ratio.
The calibration uncertainty of the radiometers is approximately 0.3~$\%$, as quoted by NIST.

We calculate the calibration factor (CF), $\frac{\int I_{\lambda}f(\lambda - \lambda') d\lambda}{i}$, 
from both the SIRCUS laser and SCL measurements, where $I_{\lambda}$ indicates the absolutes brightness of the integrating sphere viewed by the LRS,
$f(\lambda - \lambda')$ is the response function of the LRS determined from the slit width and the point spread function measured in the laboratory \citep{2013ApJS..207...33T},
and $i$ is the photocurrent measured by the LRS during the calibration measurement after correcting for the non-linearity of the detector \citep{2013ApJS..207...33T}.

Figure \ref{fig:cf} shows the CFs for the second flight after correcting for the transmittance of the additional window used in the laboratory.
The 1$\sigma$ statical uncertainty is estimated to be $<$~0.1~$\%$ from the variance across all of the detector array pixels.
The measured CFs are consistent within a 3~$\%$ r.m.s variation, which sets our systematic uncertainty of the surface brightness calibration.

\subsection{Flat Field Correction}\label{sec:flat}

\begin{figure}[htbp]
\epsscale{0.5}
\plotone{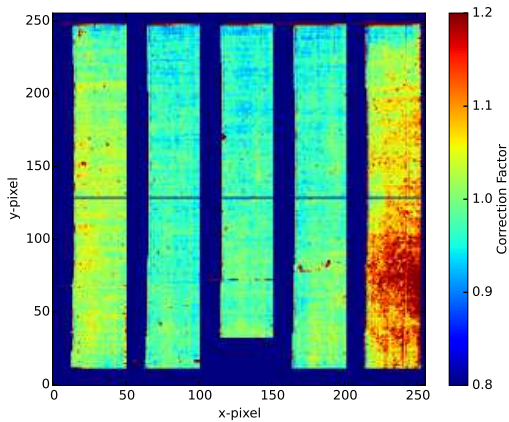}
\caption{Flat-field correction map of the second flight. }
\label{fig:flat}
\end{figure}

Under a given level of illumination, the pixel-to-pixel gain variation in the detector response generates artificial spatial fluctuations. 
The responsivity (flat-field) correction is critical to measuring spatial fluctuations in the DGL brightness.
We measure the LRS flat-field response in the laboratory using the same set-up as described Subsection \ref{sec:sb_cal}.

The measurement is made with two different integrating spheres, one with a 4 inch exit port and the other with an 8 inch port.
The spheres are filled with light from three different light sources; the SIRCUS laser, the SCL and a quartz-halogen lamp with a solar-like filter \citep{2013ApJS..207...31Z}.
The variety of source-sphere combinations enables us to check the systematic uncertainty of the flat-field correction.
The illumination pattern of the output from these integrating spheres is uniform to better than 1$\%$ \citep{2013ApJS..207...34K}, 
which is good tracer for the flat-field correction. 

Figure \ref{fig:flat} shows the flat-field correction map.
We calculated a flat fiel, $f = I_{\lambda, {\rm mean}}/I_{\lambda, {\rm pixel}}$, where $I_{\lambda, {\rm pixel}}$ indicates the brightness of the light sources detected by a pixel and 
$I_{\lambda, {\rm mean}}$ indicates the mean brightness of the light sources, for all pixels.
Typically, the flat-field correction of $f$~=~0.97-1.03~$\%$ applied to the image with an accuracy $\sim$~3~$\%$.
The accuracy of the flat-field is estimated from the variation between measurements with different set ups.

\section{Data Reduction}

\subsection{Image from raw data}
The first step of the data reduction is to make an image from individual time-ordered array reads.
For a charge integrating detector, the time derivative of the charge is proportional to the optical power. 
We fit a slope and offset to the raw data with a least squares method \citep{1993SPIE.1946..395G}, to derive the best-fit slope in photocurrent units (e-~s$^{-1}$).
A spectral image of the NEP field from the second flight is shown in Figure \ref{fig:img} as an example. 
The five vertical lines correspond to the images of five slits, which are dispersed along the horizontal direction by the prism. 
Bright stars seen as bright horizontal lines are masked to obtain the diffuse signals only.
The dark regions located at the top, bottom and left side are regions masked from infrared light.

\subsection{Removal of Bright Stars}
It is necessary to remove bright point sources from the image to isolate the diffuse components.
We first average the photocurrent of each slit along the horizontal direction,
then clip the pixels containing stars determined by the criterion that the band-averaged photocurrent is larger than the mean band-averaged photocurrent of all pixels by 2$\sigma$, 
where $\sigma$ is the standard deviation of the photocurrent. 
We iterate this clipping procedure until the ratio of the number of rejected pixels to remaining pixels is less than 0.1 $\%$ of the total. 
An example masked image is shown in the right panel of Figure \ref{fig:img}. 
All of the stars brighter than 13th magnitude are removed by this procedure, and 95 $\%$ of 13th magnitude stars are also removed.

\subsection{Dark Current} \label{dakr-current}
In the absence of incident photons, the detector produces a small positive signal called ``dark current''. 
In order to measure the absolute spectrum of the astrophysical sky, an accurate subtraction of the dark current is required.

We estimate the dark current from the masked region of the array for each observation.
The dark current is typically $\sim$~1 e-~s$^{-1}$, corresponding to $\sim$~20~nW~m$^{-2}$~sr$^{-1}$ at 1.25~$\mu$m. 
The dark current of the masked region is slightly different from that of the FOV region, based on shutter-closed data.
This difference makes a $\sim$~0.03~e-~s$^{-1}$ systematic offset corresponding to 0.7~nW~m$^{-2}$~sr$^{-1}$.
As the DGL determination uses a fit to a spatial template, residual dark current adds a small amount of noise but does not bias the result as described in Section \ref{seq:sys}.

\section{Data Analysis}\label{sec:data_analysis}

\begin{figure}[htbp]
\epsscale{1}
\plotone{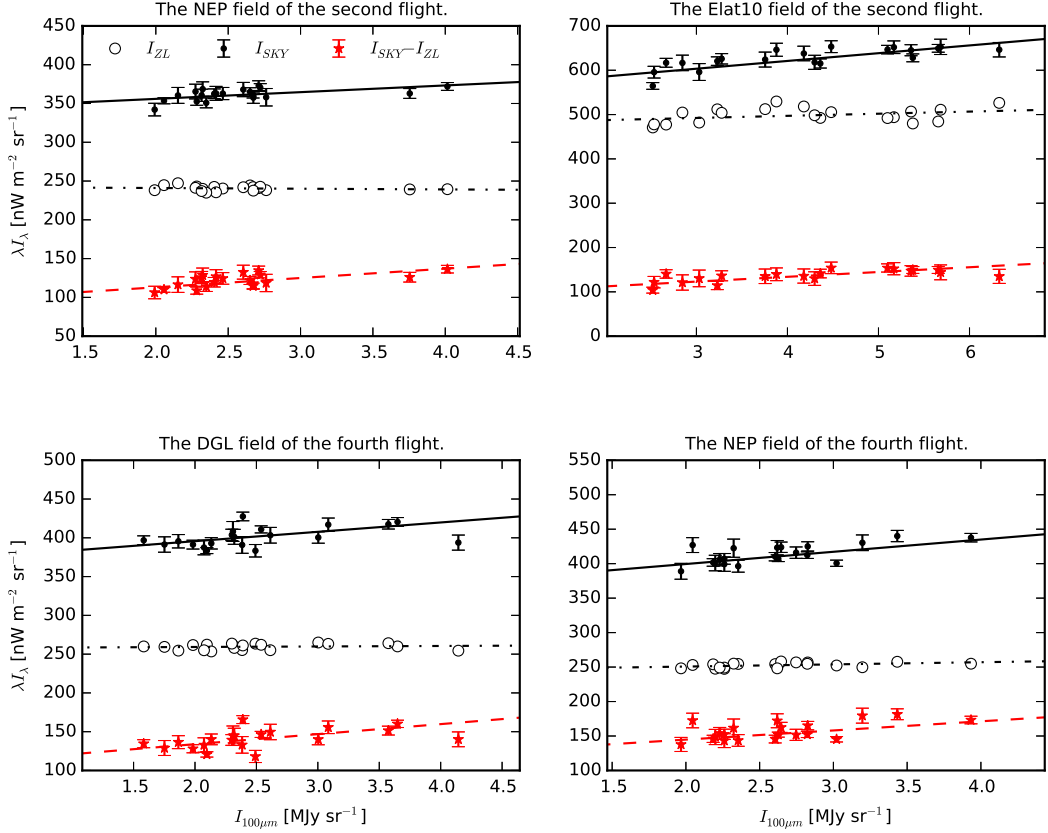}
\caption{The correlation of near-infrared surface brightness at 1.25~$\mu$m and 100~$\mu$m brightness.
         Each panel shows a different field.
         There is a statistically significant correlation between near-infrared surface brightness and 100~$\mu$m brightness in only these four fields.
         The small circles present the sky brightness, $I_{\rm sky}$, with a best-fit line black solid line. 
         The large circles indicate the subtracted ZL, $I_{\rm ZL}$, which is estimated from a ZL model \citep{1998ApJ...508...44K}, fitted as a dot-dashed line. 
         The red asterisks indicates $I_{\rm sky} - I_{\rm ZL}$, which consists of DGL, ISL, and EBL, and is fitted by the dashed line.} 
\label{fig:all_field}
\end{figure}

\begin{figure}[htbp]
\epsscale{1}
\plotone{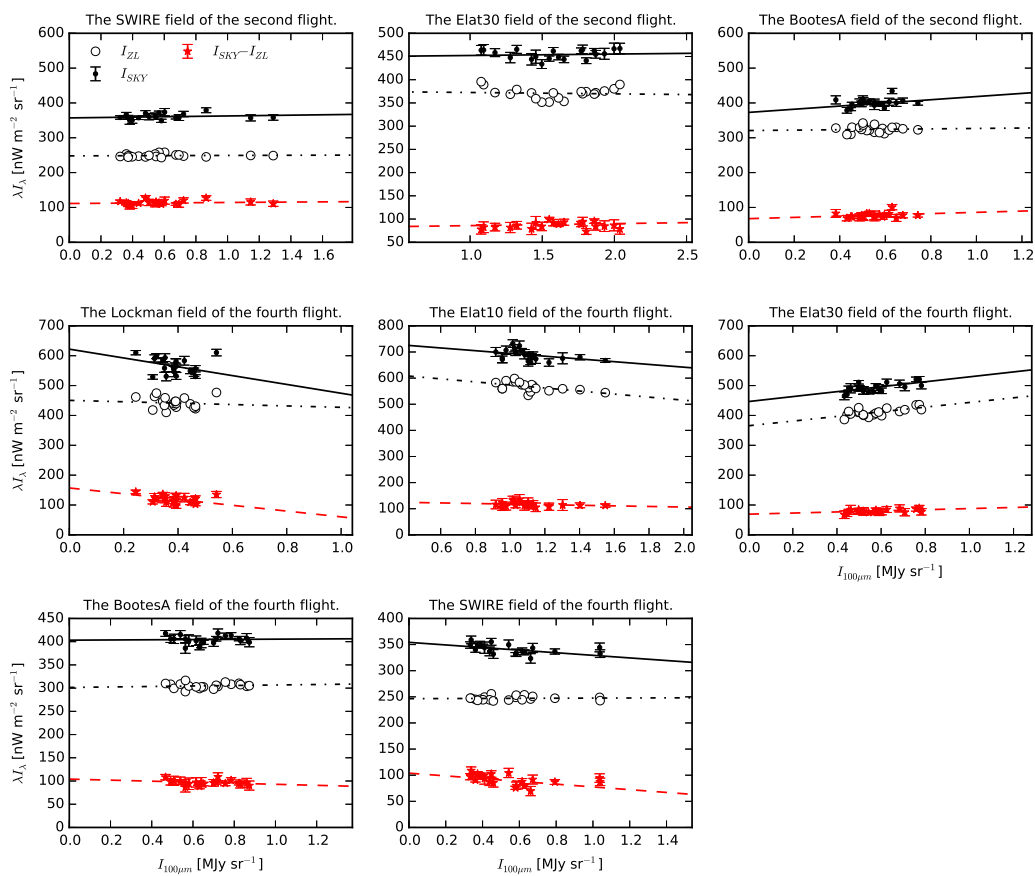}
\caption{Because the contrast of the DGL brightness is too low in these fields, there is no statistically significant correlation between near-infrared surface brightness and 100~$\mu$m brightness. 
See Figure \ref{fig:all_field} caption.
	      } 
\label{fig:all_field_2}
\end{figure}

\begin{figure}[htbp]
\epsscale{1}
\plotone{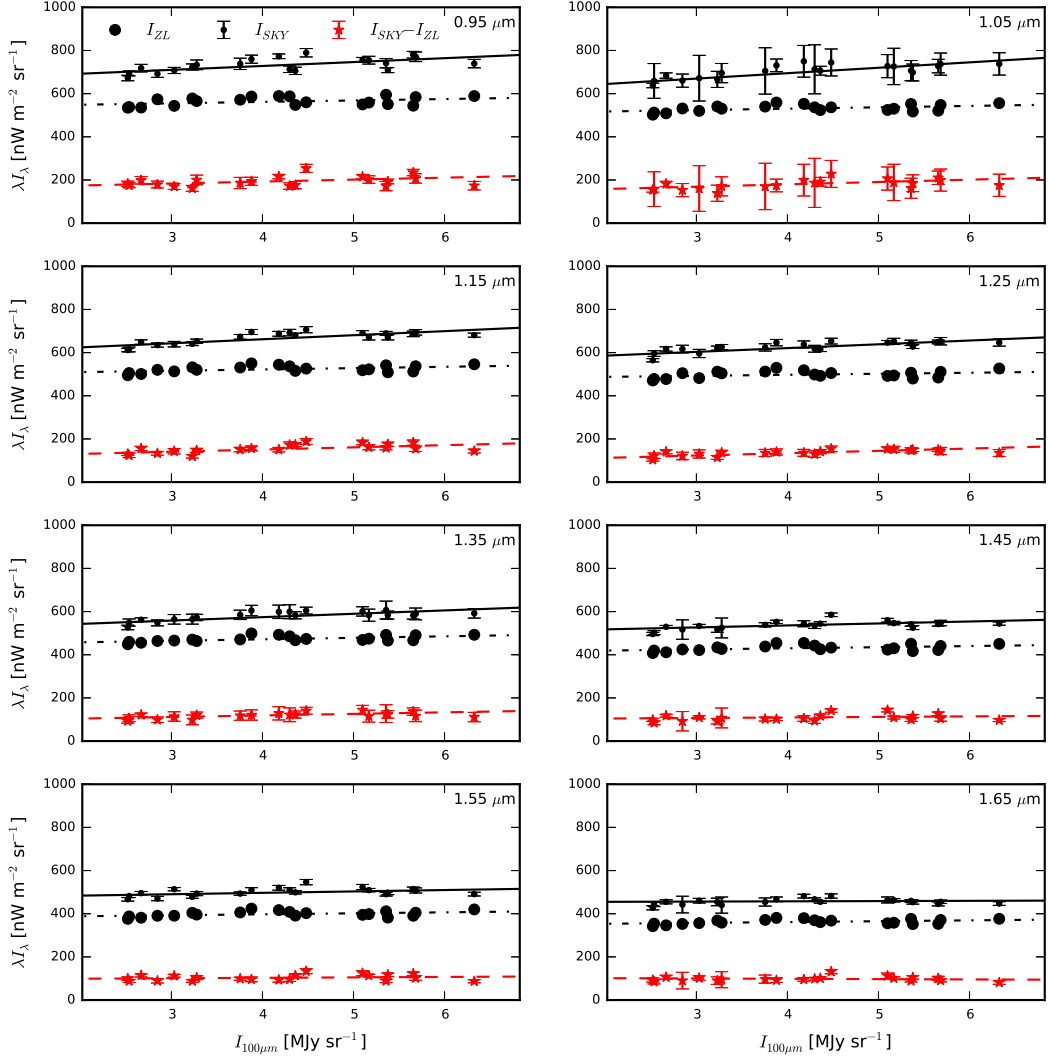}
\caption{The correlation of near-infrared and 100~$\mu$m brightness in the Elat10 field observed in the second flight.
         Each panel shows a different observed wavelength.
         See Figure \ref{fig:all_field} caption. }
\label{fig:corr_e10}
\end{figure}

\begin{figure}[htbp]
\epsscale{1}
\plotone{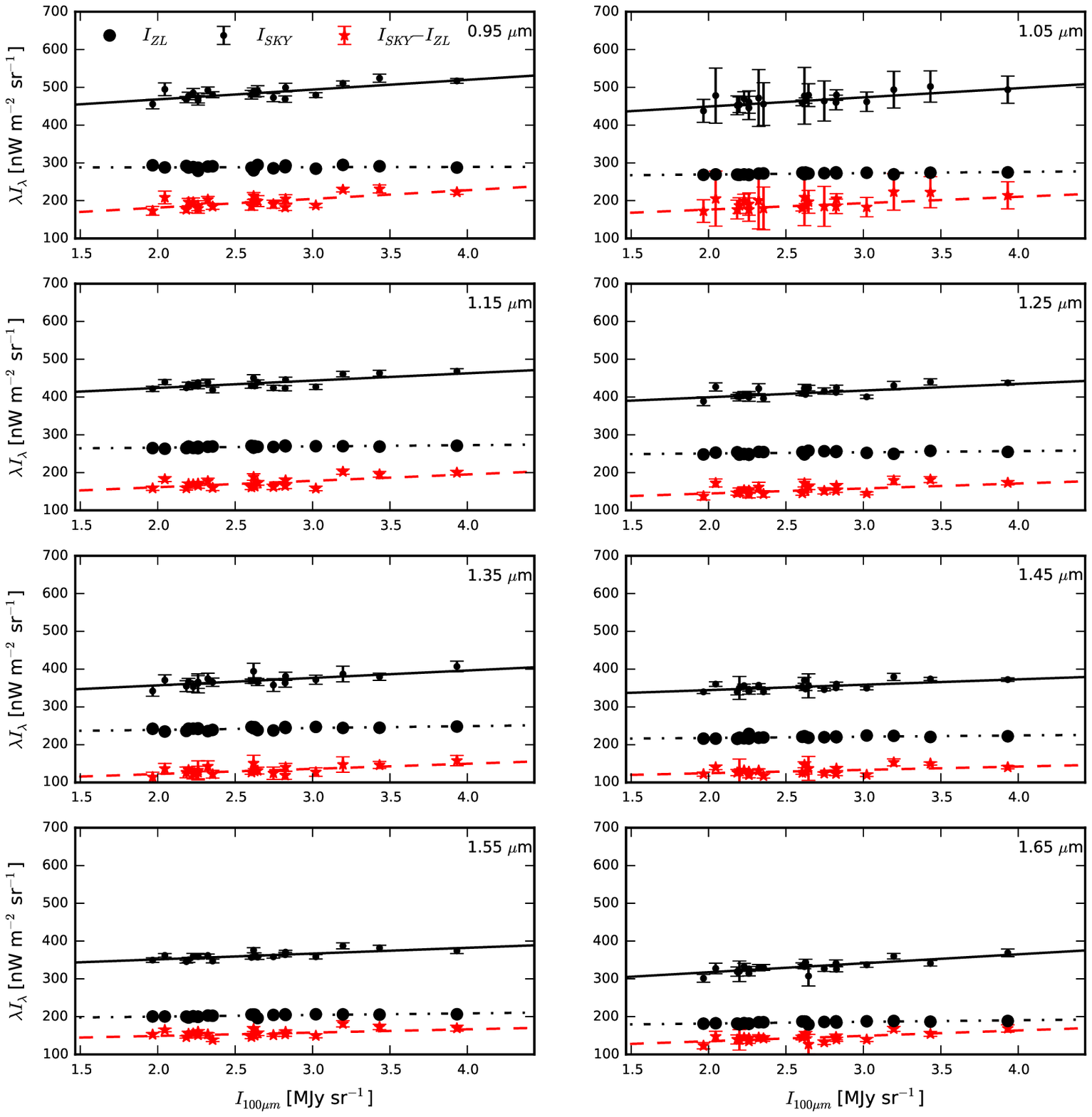}
\caption{Correlation in the NEP field observed in the fourth flight. See Figure \ref{fig:all_field} caption. } 
\label{fig:corr_nep}
\end{figure}

The processed data includes astronomical emission from ZL, DGL, ISL and EBL, {\it i.e.},
\begin{equation}
  I_{{\rm sky}} = I_{{\rm ZL}} + I_{{\rm DGL}} + I_{{\rm ISL}} + I_{{\rm EBL}}.
\end{equation}
We separate the DGL component using its spatial distribution as traced by 100~$\mu$m emission with
structure on spatial scales smaller than a degree as shown in Figure \ref{fig:100um_map_2nd} and  \ref{fig:100um_map_4th}.
ZL is known to be spatially uniform on spatial scales smaller than a degree \citep{1997A&A...328..702A, 2005Natur.438...45K, 2012ApJ...760..102P}. 
In this analysis, we assume that there is no correlation between the EBL and the 100 $\mu$m brightness.

Although the small-scale fluctuations of the ZL brightness are negligible, the large-scale distribution may affect the correlation between the 100 $\mu$m brightness and the DGL brightness.
To account for contamination from the ZL, LRS spectral images are constructed by aligning the DIRBE-based model predictions
for large scale ZL structure with the location of the LRS slits. 
We use two different models to check the systematic effect of the ZL subtraction \citep{1998ApJ...508...44K, 1998ApJ...496....1W}.

We first obtain a fiducial ZL spectrum by differencing the LRS data between fields, i.e, $(I_{{\rm sky}, i} - I_{{\rm ISL}, i}) - (I_{{\rm sky}, j} - I_{{\rm ISL}, j})$.
The difference is approximately $I_{ZL,i} - I_{ZL, j}$ when the DGL brightness is similar in field $i$ and $j$, 
where $I_{ZL, i}$ indicates the ZL spectrum, and $I_{sky, i}$ indicates the sky spectrum.
The ISL spectrum, $I_{{\rm ISL}, i}$, is estimated by Monte-Carlo simulations of the star distribution in the FOV using the 2MASS catalogue \citep{2006AJ....131.1163S} 
and a population synthesis code for simulating Galactic star counts \citep{2005A&A...436..895G},
taking into account the limiting magnitude and effective slit area of the LRS \citep[in prep]{2014matuura}.
We calculate the difference of every combination of fields and take an average of the differences to make the ZL spectrum template.
Because the difference of every combination is consistent with each other within the errors , we use this ZL spectrum template for every field.
The absolute brightness of ZL is estimated by using ZL models for the DIRBE data at 1.25~$\mu$m \citep{1998ApJ...508...44K, 1998ApJ...496....1W}.


We separate the DGL component from $I_{\rm sky} - I_{\rm ZL}$ using a linear correlation analysis.
In the optically thin limit, the DGL brightness can be approximated to be:
\begin{equation}
  I_{DGL} \approx \gamma_{NIR} \sigma_{NIR} N I_{ISRF}, 
\end{equation}
where $\gamma_{NIR}$ and $\sigma_{NIR}$ present albedo and scattering cross sections of interstellar dust in the near-infrared respectively, $I_{ISRF}$ is the brightness of the ISRF,  and $N$ is column density of interstellar dust.
In the optically thin limit, the far-infrared brightness is given by:
\begin{equation}
 I_{FIR}  \approx (1 - \gamma_{FIR}) \sigma_{FIR}N B(T),
\end{equation}
where the $B(T)$ is Planck function. 
We assume the temperature of interstellar dust is uniform,
so that the DGL brightness can be written as a function of the far-infrared brightness;
\begin{equation}
 I_{DGL}  \approx \frac{\gamma_{NIR} \sigma_{NIR} I_{ISRF}}{(1 - \gamma_{FIR})\sigma_{FIR} B(T)} I_{FIR}.
 \label{eq:dgl_linear}
\end{equation}
Equation \ref{eq:dgl_linear} indicates that the DGL brightness correlates linearly with the far-infrared brightness in optically thin regions.

Because thermal emission from interstellar dust dominates the diffuse sky brightness at 100~$\mu$m, 
it is easier to separate interstellar radiation from other diffuse sources in the far-infrared than in the near-infrared. 
As a result, we use a far-infrared intensity map measured at 100~$\mu$m \citep{1998ApJ...500..525S}.
The SFD 100~$\mu$m map is based on the all-sky survey combination of IRAS (IR Astronomical Satellite) and DIRBE/COBE (Diffuse Infrared Background Experiment / Cosmic Background Explorer). 
The SFD 100~$\mu$m map has the accurate calibration of DIRBE/COBE and the $\sim$~6$'$ resolution of IRAS .

Thus the linear correlation between our data and 100~$\mu$m brightness is written as:
\begin{equation}
 I_{\rm sky} - I_{\rm ZL} = a(\lambda) + b(\lambda) I_{100\mu m}.
\label{eq:linear_corr}
\end{equation}
The slope $b(\lambda)$ gives the conversion factor from 100~$\mu$m brightness to DGL brightness. 
The offset $a(\lambda)$ accounts for the ISL and EBL contributions.

\section{Results}

Figures \ref{fig:all_field} and \ref{fig:all_field_2} show the correlation between $I_{\rm sky} - I_{\rm ZL}$ with the brightness of the 100~$\mu$m map in all fields at 1.25~$\mu$m. 
There is a statistically significant correlation in only four fields with large spatial contrast in 100~$\mu$m brightness, specifically the NEP field of the second and fourth flight, the DGL field, and the Elat10 field of the second flight.
The Peason correlation coefficient, 
\begin{equation}
 r = \frac{\sum_{k = 1}^{N_{\rm data}}(I_{{100\mu {\rm m}}, k} - \overline{I}_{100\mu {\rm m}})(I_{{\rm sky}, k} - \overline{I}_{\rm sky} - I_{{\rm ZL}, k} - \overline{I}_{\rm ZL})} {\sqrt{\sum_{k = 1}^{N_{\rm data}}(I_{{100\mu {\rm m}}, k} - \overline{I}_{100\mu {\rm m}})^{2}} \sqrt{\sum_{k = 1}^{N_{\rm data}}(I_{{\rm sky}, k} - \overline{I}_{\rm sky} - I_{{\rm ZL}, k} - \overline{I}_{\rm ZL})^{2}} },
\end{equation}
where $N_{\rm data}$ indicates the number of data and $\overline{I}$ indicates the average of the data,
is larger than 0.5 at $\lambda$~$<$~1.35$\mu$m only in these four fields.
The Peason correlation coefficients of other fields are less than 0.5, indicating that 
the contrast of the DGL brightness is too low to measure DGL in other fields.
Thus we use the results of the four statistically significant fields below. 
Figures \ref{fig:corr_e10} and \ref{fig:corr_nep} show the correlation of $I_{\rm sky} - I_{\rm ZL}$ with 100~$\mu$m brightness for the NEP field and the Elat10 field of the second flight at different wavelengths.
We also present $I_{\rm ZL}$, estimated from the ZL model and $I_{\rm sky}$ in Figures \ref{fig:all_field}, \ref{fig:all_field_2}, \ref{fig:corr_e10} and \ref{fig:corr_nep}.
To improve S/N, array pixels are binned along the slit length direction and averaged into broad $\Delta \lambda$ = 200 nm wavelength bins.
The error bars present 1 $\sigma$ statistical uncertainty estimated from the variance across all pixels in the binned region, added in quadrature with the $\sim$3~$\%$ flat-field error.

As shown in Figure \ref{fig:corr_e10}, the scatter of $I_{\rm sky}$ is traced by $I_{\rm ZL}$, and the scatter of $I_{\rm sky} - I_{\rm ZL}$ is smaller than that of $I_{\rm sky}$.
This result indicats that a large-scale ZL gradient is present, and that this gradient is explained by the ZL model. 

\begin{figure}[htbp]
\epsscale{1.0}
\plotone{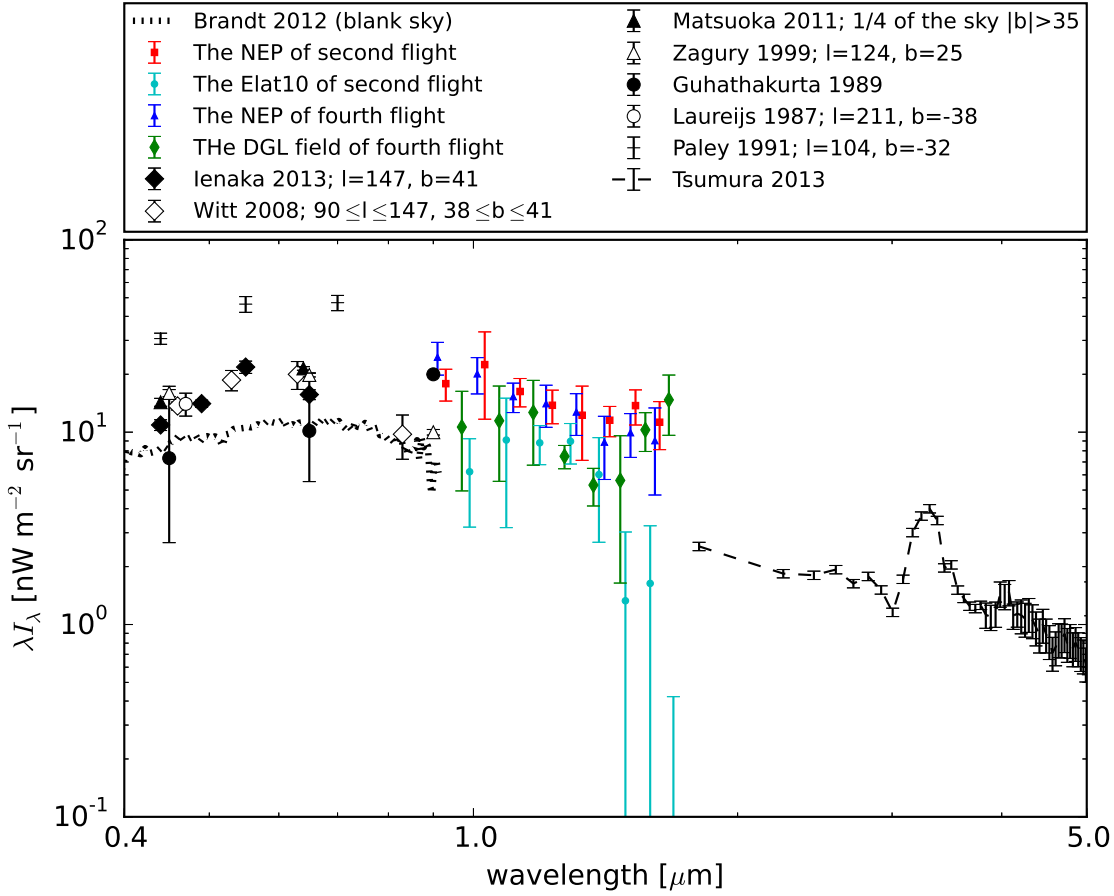}
\caption{The DGL spectrum of all four fields. The intensity is normalized at $I_{100\mu m}$ = 1 MJysr$^{-1}$. The red square indicates the DGL spectrum of the NEP field of the second flight, the cyan small circles the Elat10 field,
the green diamonds the DGL field, and the blue triangles the NEP field of fourth flight. The error bars show 1 $\sigma$ standard error.
The optical measurements are also presented.
\citet{1987A&A...184..269L, 1989ApJ...346..773G, 1999A&A...352..645Z} and \citet{ 1991ApJ...376..335P} measure the DGL brightness correlating with the optical brightness with the original IRAS 100~$\mu$m map. 
Recent optical measurements (\citet{2008ApJ...679..497W, 2011ApJ...736..119M, 2013ApJ...767...80I}) correlate the optical brightness with the SFD 100~$\mu$m map.
Only \citet{2012ApJ...744..129B} measures the spectrum of DGL at the optical wavelengths.
\citet{2013PASJ...65..120T} measures the spectrum of DGL at 1.8~$\sim$~5.0~$\mu$m and detects a polycyclic aromatic hydrocarbon feature.
}
\label{fig:dgl_all_field}
\end{figure}

In Figures \ref{fig:all_field}, \ref{fig:all_field_2}, \ref{fig:corr_nep} and \ref{fig:corr_e10}, we fit the linear function from Equation \ref{eq:linear_corr} to our data, and
Figure \ref{fig:dgl_all_field} shows the slope $b(\lambda)$ as a function of wavelength. 
The results of the four fields are shown in Figure \ref{fig:dgl_all_field} with the results of various optical and infrared measurements.

The DGL spectrum of the NEP field observed in the second flight is consistent with that of the fourth flight within our errors.
We obtain the same result from observations with different calibration measurements, which indicats that our measurements and calibrations are valid.
The DGL spectrum of the Elat10 field is systematically low compared to the NEP field.

Figure \ref{fig:dgl_spec} shows the mean DGL spectrum of the four fields.
DGL measurements at optical wavelengths indicate that the DGL brightness varies from cloud to cloud by a factor of 3-4.
\citet{2013ApJ...767...80I} argued that variation of optical depth and/or the forward scattering characteristics of dust grains might be the origin of this variation.
Because we measured the DGL brightness in diffuse regions, away from dense molecular clouds and star forming regions, we only present optical measurements in similarly diffuse regions for comparison. 
However, because the DGL spectrum of BD12 had a large uncertainty in flux calibration, we do not present its spectrum. 
The Pioneer data \citep{2011ApJ...736..119M} are presented at optical wavelengths, since the data covered one fourth of sky at galactic latitude higher than 35$^\circ$.
As shown in Figure \ref{fig:dgl_spec}, our result is consistent with interpolated levels from the Pioneer data and the AKARI data \citep{2013PASJ...65..119T}.

\begin{figure}[htbp]
\epsscale{1.0}
\plotone{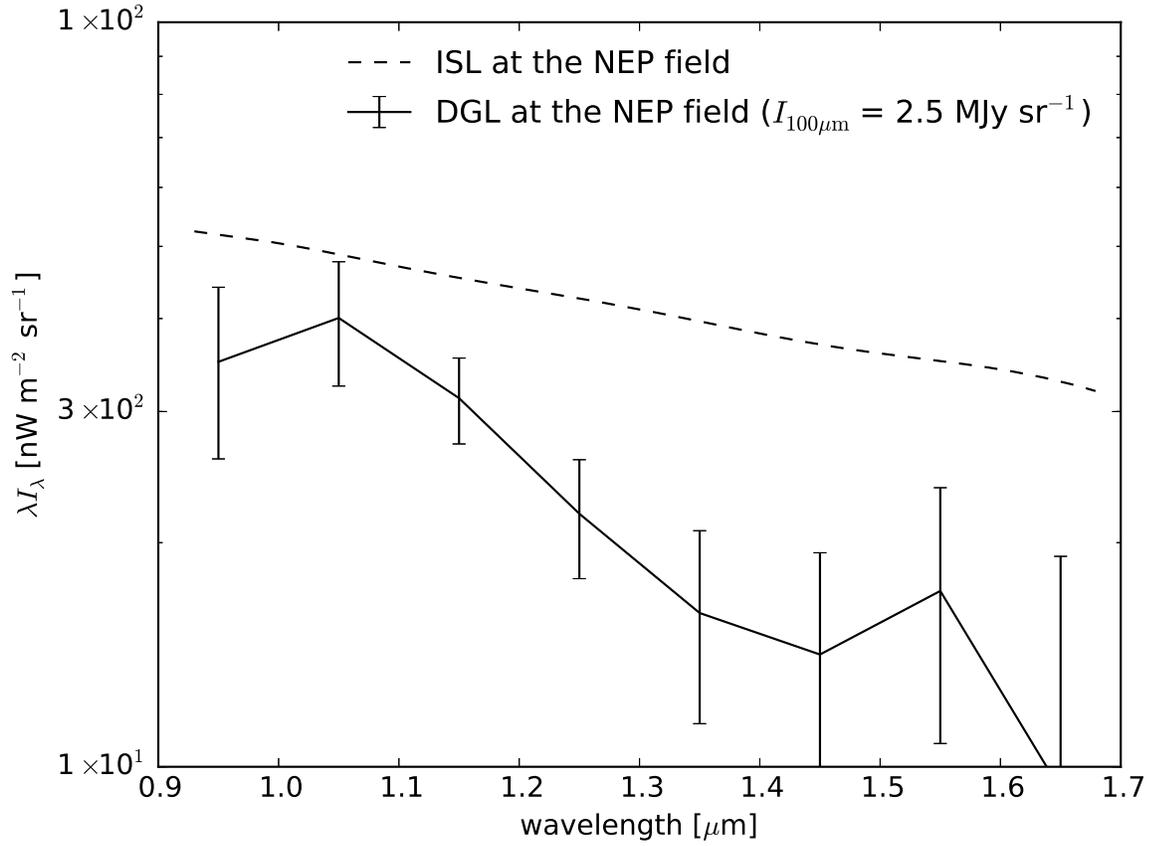}
\caption{Comparison between the mean DGL spectrum of all four fields, given by the solid line,  and the ISL spectrum, given by the dashed line for stars fainter than 13th magnitude.
	     The error bars give 1 $\sigma$ statistical error. }
\label{fig:dgl_isl}
\end{figure}

\begin{figure}[htbp]
\epsscale{1.0}
\plotone{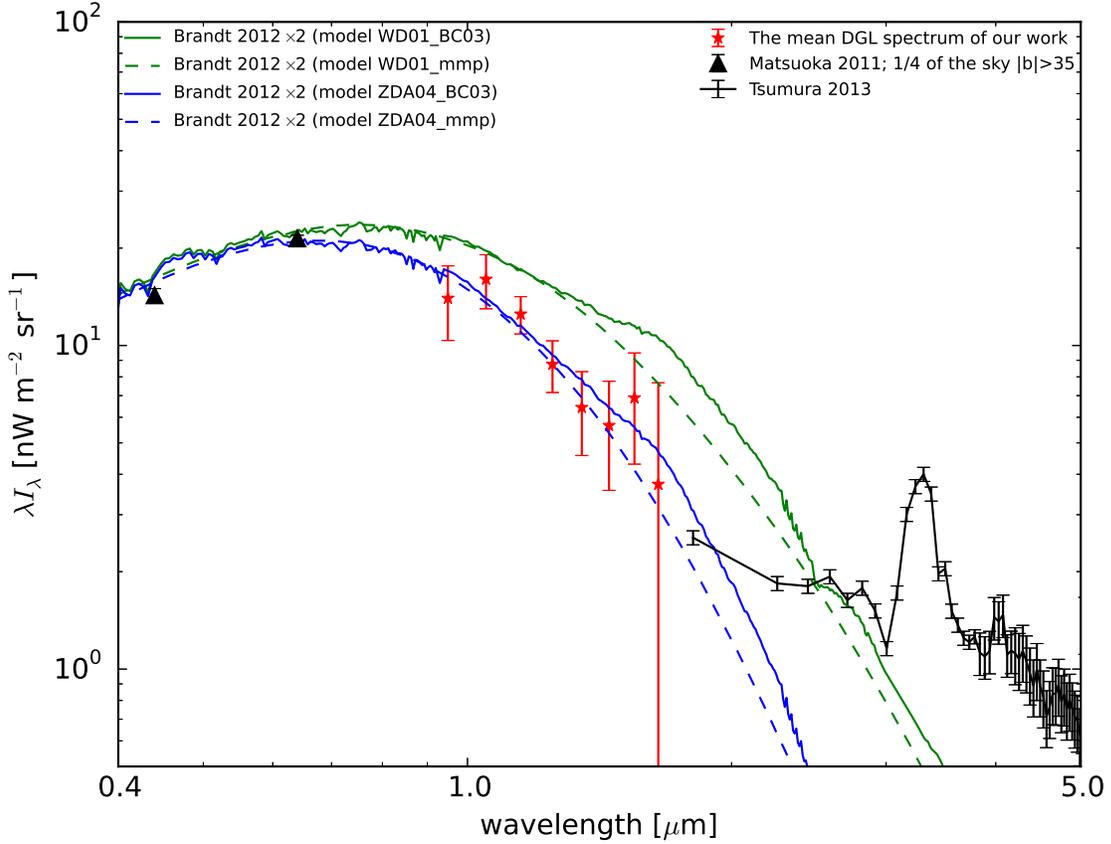}
\caption{Comparison between the mean DGL spectrum and theoretical DGL models described in Table \ref{tbl:brandt_model}. The 100~$\mu$m intensity is normalized to that of 1~MJy~sr$^{-1}$. 
The red asterisks indicate our mean DGL spectrum given in Table \ref{tbl:brandt_model}. 
We only show the DGL brightness measured from diffuse sky regions \citep{2011ApJ...736..119M, 2013PASJ...65..120T} to compare with our results.}
\label{fig:dgl_spec}
\end{figure}

\section{Systematic Uncertainty}\label{seq:sys}

\begin{figure}[htbp]
\epsscale{1.0}
\plotone{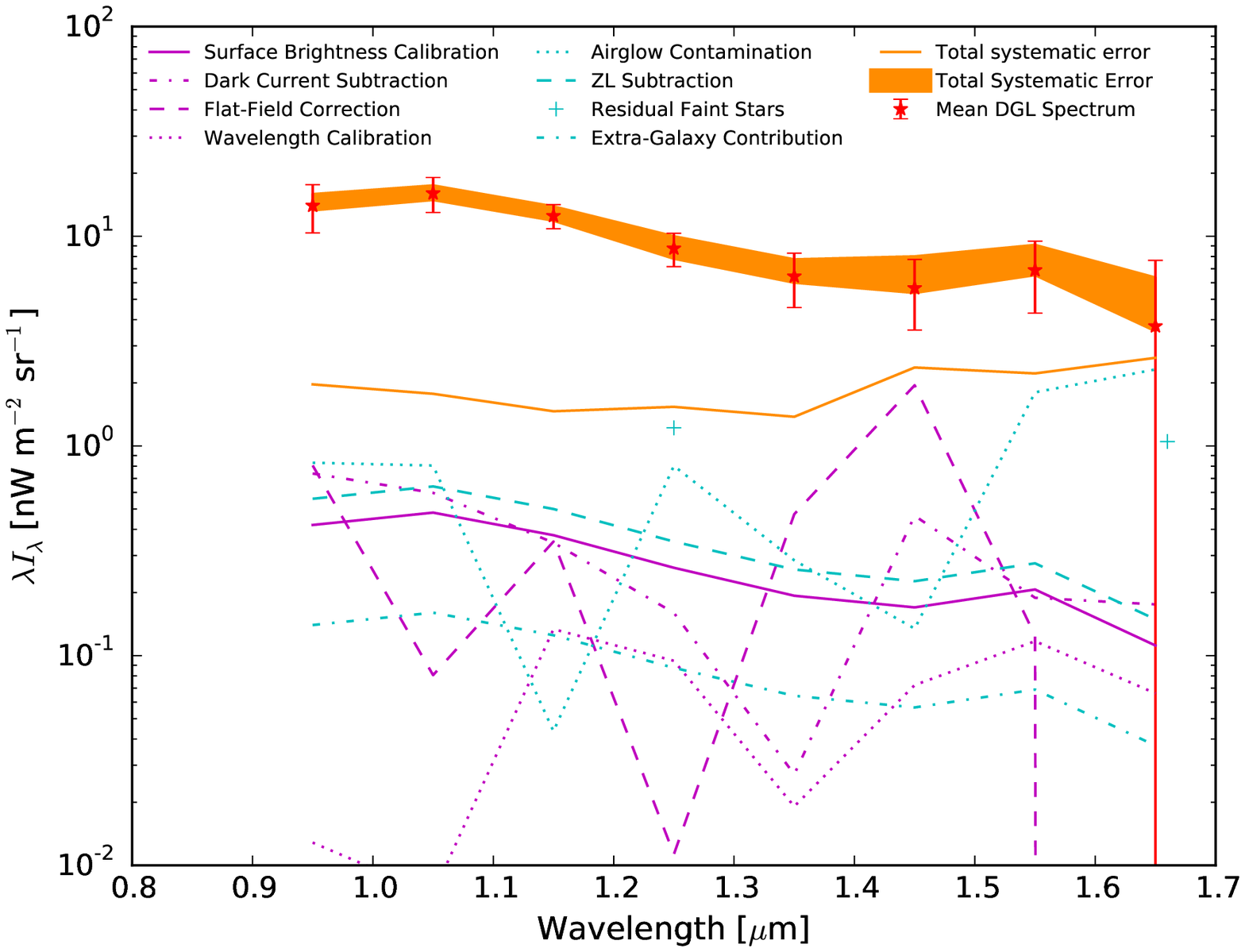}
\caption{The systematic errors on the mean DGL spectrum.
The intensity is normalized at $I_{100\mu m}$ = 1 MJysr$^{-1}$.
The instrumental systematic errors (purple line) associated with surface brightness calibration error (solid line), 
dark current subtraction (long dashed dotted line), 
flat-field error (dashed line),
and wavelength calibration error (dotted line) are indicated.
The systematic error from the astronomical foreground (cyan line) associated with airglow contamination (dotted line),
ZL subtraction (dashed line),
and residual faint stars (dotted line) are presented.
The orange solid line indicates the total instrument and astronomical systematic error that is the quadrature sum of the systematic errors.
The mean DGL spectrum is presented as the red asterisks with 1$\sigma$ statistical error.
An orange shaded region shows the total systematic error band about the mean DGL spectrum.}
\label{fig:airglow_isl_corr}
\end{figure}

We estimate the possible systematic error from residual airglow emission, residual faint stars, and uncertainties from the ZL subtraction.

\subsection{Wavelength Calibration}
We estimate the systematic error from the wavelength calibration by shifting the wavelength map by $\pm$ 1~nm, 
which corresponds to the systematic error seen in multiple wavelength calibration measurements.
We derive the DGL spectra using this shifted wavelength map and calculate the differences of the DGL spectra.
As shown in Figure \ref{fig:airglow_isl_corr}, this mean difference on the mean DGL spectrum is negligible.

\subsection{Flat-Field Correction}
We bound the systematic error of the flat-field correction based on the difference between the flat-field measurements using different integrating spheres.
We use two integrating spheres with 4" and 8" exit port diameters as described in Subsection \ref{sec:flat}.
We apply each flat-field and derive the DGL spectra in four fields.
We then calculate the difference between the DGL spectra of each flat-field and calculate mean of this difference.
We determine the mean difference as the systematic error on the mean DGL spectrum of the flat-field correction, as shown in Figure \ref{fig:airglow_isl_corr}.
The flat-field systematic error is $<$~6~$\%$, which is acceptably small.

\subsection{Dark Current Subtraction}
To check the systematic error from the dark current subtraction, 
we use the shutter closed data acquired during the flights.
We subtract the dark current from the shutter closed data using the same method for the field data described in \ref{dakr-current}, 
and then correlate with 100~$\mu$m brightness.
There is no correlation detected between the shutter closed data and the 100~$\mu$m brightness; the Pearson correlation coefficient is less than 0.3 in all wavelengths at all fields.
The mean slope of this correlation of four fields is determined to be the worst-case systematic error of the dark current subtraction and is shown in Figure \ref{fig:airglow_isl_corr} as a function of wavelength.

\subsection{Airglow}
To estimate the contamination from airglow, we extracte the spatial structure of the airglow emission by differencing the first-half and the second-half integration of each field.
We attribute a time and altitude dependence to the airglow emission, so the observed brightness is written as
\begin{equation}
I_{{\rm obs}}(t, h) = I_{{\rm sky}} + I_{{\rm air}}(t, h)
\label{eq:airglow}
\end{equation}
where $I_{air}$ indicates the brightness of airglow emission, $t$ indicates time from the launch, and $h$ is altitude.
We make an image of the spatial structure of airglow by differencing the first-half and the second-half integration, 
$I_{\rm obs}(t_{\rm first}, h_{\rm first}) - I_{\rm obs}(t_{\rm second}, h_{\rm second}) =  I_{\rm air}(t_{\rm first}, h_{\rm first}) - I_{\rm air}(t_{\rm second}, h_{\rm second})$, to cancel the astronomical component.

To estimate the systematic contribution of the airglow emission to the DGL spectrum, we correlate this airglow image with the 100~$\mu$m brightness map. 
The Pearson correlation coefficient is less than 0.5 in all wavelengths at all fields.
This result indicates that there is no correlation between the airglow emission and 100~$\mu$m brightness.
From these results, we conclude that the airglow emission does not have a systematic effect on the DGL measurement.
We estimate the maximum possible systematic error of airglow on the mean DGL spectrum from the mean slope of this correlation between airglow image with the 100~$\mu$m brightness map of four fields,
as shown in Figure \ref{fig:airglow_isl_corr}.

\subsection{Zodiacal Light Subtraction}
The systematic uncertainty of the ZL subtraction is quantified by calculating the difference between the DGL spectrum derived using two different ZL models for the COBE data \citep{1998ApJ...508...44K, 1998ApJ...496....1W}.
In the Elat10 field, because the direction of the large-scale ZL gradient is similar to the spatial structure of DGL as shown in Figure \ref{fig:corr_e10}, 
the systematic uncertainty of the ZL subtraction is 10~$\sim$~30~$\%$ of the DGL brightness, depending on the wavelength.
In the other four fields, the systematic uncertainty of the ZL subtraction is $\sim$1~$\%$ of the DGL brightness.
The systematic error of the ZL subtraction on the mean DGL spectrum is presented in Figure \ref{fig:airglow_isl_corr}.

\subsection{Residual Faint Stars}
Residual faint stars may spatially correlate with 100~$\mu$m brightness and make a systematic effect.
To rule this out, we correlate the integrated brightness of stars fainter than 13th magnitude from 2MASS at H- and J-bands with the 100~$\mu$m brightness map. 
Although the contribution of faint stars is brightest in the NEP fields,
the Pearson correlation coefficient of the NEP fields is $\sim$ 0.3 at H and J-bands.
This indicates that there is no correlation between the faint stars with the 100~$\mu$m brightness.
There are also no correlations in the DGL and Elat10 fields.
We determine the worst-case systematic error on the mean DGL spectrum as the mean slope of this correlation between faint stars and 100~$\mu$m brightness map of four fields, as shown in Figure \ref{fig:airglow_isl_corr}.

\subsection{Contribution of Extra-Galactic Background Light}
According to \citet{1998ApJ...500..525S}, galaxies with fluxes brighter 1.2 Jy were removed when they construct the SFD~100~$\mu$m map. 
Galaxies fainter than this limit have a peak flux density of $\sim$~0.12~MJy sr$^{-1}$ at 100~$\mu$m. 
\citet{1998ApJ...500..525S}, however, mentioned that the contamination from extragalactic objects was very nearly uniformly distributed. 
\citet{2011ApJ...737....2M} measured the far-infrared sky by AKARI at 90~$\mu$m and estimated that the spatial fluctuation due to galaxies was $<$~0.01~MJy sr$^{-1}$ at 1$^{\circ}$ scale before removing bright galaxies. 
Thus galaxies contributed $<$~1~$\%$ of the DGL measurement, since the spatial fluctuation of the SFD 100~$\mu$m map is $>$~1~MJy sr$^{-1}$ in our fields. 
We show upper limit to the systematic error from residual galaxies on the mean DGL spectrum in Figure
\ref{fig:airglow_isl_corr}, which is negligible for our DGL measurements.

\subsection{Non-linear correlation between DGL and 100~$\mu$m brightness}
If the dust scattering is optically thick, near-infrared DGL will not correlate linearly with 100~$\mu$m brightness.
\citet{2013ApJ...767...80I} presented the linear correlation breaks at $>$~6~MJy sr$^{-1}$ in the optical wavelengths.
These results indicated that the linear correlation breaks appear at extinction $A(\lambda) > 0.5$, where $A(\lambda)$ is extinction at wavelength $\lambda$ \citep{1998ApJ...500..525S}.
Since the 100~$\mu$m brightnesses in our observed fields is always $<$~6~MJy$^{-1}$ corresponding to $A(0.9~\mu {\rm m}) = 0.4$, we expect the correlation to be linear.

\subsection{Systematic Error of the Mean DGL Spectrum}
We combine the systematic errors of the final mean DGL spectrum as shown in the orange band of Figure \ref{fig:airglow_isl_corr}.
The total systematic error is 30~$\%$ of the mean DGL spectrum at $\lambda ~ \leq$~1.35~$\mu$m, and $>$~30~$\%$ of the mean DGL spectrum at $\lambda ~ >$~1.35~$\mu$m.
The total systematic error 
is the maximum case because there is no statistically significant correlation between dark current, airglow and residual faint stars with 100~$\mu$m brightness.

\section{Discussion}

As seen in Figure \ref{fig:dgl_all_field}, the slope $b(\lambda)$ of the NEP and DGL fields are consistent.
The Elat10 field is marginally lower in amplitude, although the errors are large.
It is possible that the large ZL correction in this field is partly responsible,
or the result may be from spatial dependence of the DGL spectrum.
Figure 4 of BD12 also shows the Galactic latitude and the Galactic longitude dependence of the slope $b(\lambda)$ at optical wavelengths. 
If the difference in slope of $b(\lambda)$ between fields is real,
it could be attributed to spatial variations of dust scattering properties in the ISRF spectrum. 
However, it is difficult to claim significance from our data alone.
To definitively detect spatial variations, higher S/N observations and wider sky coverage are required.

Figure \ref{fig:dgl_isl} shows the mean spectrum of stars fainter than 13 magnitude in the NEP field,
obtained from the TRILEGAL code \citep{2005A&A...436..895G}. 
The mean DGL spectrum shows a bluer color than the ISL spectrum.
Small grains are mainly responsible for DGL production,
since scattering by large particles would not show the observed reddening.
The LRS result is consistent with measurements of polarization of starlight by interstellar dust,
which suggest the grains responsible for polarization of starlight have diameter $a \approx 0.1~\mu$m \citep{1995ApJ...444..293K}. 
This result is also consistent with observed stellar extinction from the infrared, 4~$\mu$m, to the ultraviolet, 0.1~$\mu$m \citep{2001ApJ...548..296W, 2004ApJS..152..211Z}.
WD01 and ZDA04 constructed models to reproduce the observed stellar extinction, and argued that interstellar dust is composed of grains with 10$^{-3}$~$\mu$m$~<$~$a$~$<$~1~$\mu$m peak at $a$~$\sim$~0.1~$\mu$m.

\begin{deluxetable}{crrrrrrrr}
\tabletypesize{\scriptsize}
\tablecaption{A summary of the interstellar dust scattering models of Brandt $\&$ Drain (2012), based on Weingartner $\&$ Draine (2001) (WD01), Zubko et al (2004) (ZDA04), Maths et al. (1983) (MMP83), and Bruzual $\&$ Charlot (2003) (BC03).}
\tablewidth{0pt}
\tablehead{\colhead{Model Name} & \colhead{Composition} & \colhead{Interstellar Dust Size} & \colhead{Interstellar Radiation Field} \\}
\startdata
WD01-BC03 & Graphite, silicate, and PAH & a$_{0.5}$=0.12 & Stellar population synthesis \\
WD01-MMP83 & Graphite, silicate, and PAH & a$_{0.5}$=0.12 & Solar neighborhood \\
ZDA04-BC03 & Bare graphite, bare silicate, and PAH & a$_{0.5}$=0.06$\sim$0.07 & Stellar population synthesis  \\
ZDA04-MMP83 & Bare graphite, bare silicate, and PAH & a$_{0.5}$=0.06$\sim$0.07 &  Solar neighborhood\\
\enddata
\label{tbl:brandt_model}
\end{deluxetable}

To constrain the size distribution of dust particles from the LRS measurements, 
we compare the mean DGL spectrum with theoretical models from BD12 in Figure \ref{fig:dgl_spec}. 
BD12 considered an infinite plane-parallel Galaxy with a Gaussian vertical distribution of dust from $\sigma=25$pc \citep{1995ApJ...448..138M, 2003PASJ...55..191N}, 
and a two-exponential distribution of stars with scale heights of 300pc and 1350pc \citep{1998gaas.book.....B, 1983MNRAS.202.1025G}. 
BD12 estimated the stellar emission spectrum in two ways:
(1) a model that reproduces the local ISRF of \citet{1983A&A...128..212M} (hereafter MMP83), 
and (2) a stellar population synthesis model of \citet{2003MNRAS.344.1000B} (hereafter BC03) with solar metallicity and an exponential star formation history over 12Gyr. 
BC12 also used two typical dust models from WD01 and ZDA04. 
The dust of the WD01 model consisted of graphite, silicate, and PAH material,
while the dust composition of the ZDA04 model consisted of bare graphite grains, bare silicate grains and PAHs. 
The size distribution of the ZDA04 model was shifted to smaller grains compared with WD01 model.
The half-mass grain radius was $a_{0.5} \approx 0.12$~$\mu$m for both silicate and carbonaceous grains in the WD01 model, 
and $a_{0.5} \approx 0.06$~$\mu$m and $0.07$~$\mu$m for carbonaceous grains and silicate grains respectively \citep{2011piim.book.....D}. 
Dust grains of size $a \geq 0.2$~$\mu$m were absent in the ZDA04 model. We summarize these models in Table \ref{tbl:brandt_model}. 
Because the models underestimate $b(\lambda)$ by a factor of 2,
the model is normalized at 0.44~$\mu$m to the data of \citet{2011ApJ...736..119M}. 
\citet{2013ApJ...767...80I} presented two possible explanations for this;
a deficiency in UV photons in the ISRF, or that the dust grain albedo is different higher than what is assumed in the models.

The combination of our result, the Pioneer data, and the shortest band of the AKARI data can be reproduced by both models of ZDA04.
On the other hand, the models of WD01 cannot reproduce the observed DGL spectrum.
The models do not take into account PAH emission and thermal emission, so they do not reproduce the DGL spectrum measured by AKARI.
Because the effect of different interstellar radiation fields is not significant at near-infrared wavelengths, 
our result implies that interstellar dust is dominated by small particles with $a_{0.5} \approx 0.06~\mu$m, with a few large grains $a > 0.2~\mu$m.

\section{Summary}
We measure the spectrum of diffuse galactic light in the near-infrared,
helping to determine the properties of interstellar dust, particularly its size distribution. 
Since airglow emission is too bright to measure the DGL spectrum from the ground, the spectrum has never been observed at diffuse sky regions in the near-infrared. 
To derive the DGL spectrum as shown in Figure \ref{fig:dgl_all_field}, we correlate spectral images measured by the LRS with 100~$\mu$m intensity.
The measured DGL spectrum shows no resolved spectral features and is smoothly connected to the other DGL measurements in the optical and near-infrared wavelengths longer than 1.8~$\mu$m, as shown in Figure \ref{fig:dgl_spec}.
Rayleigh scattering of starlight by small grains largely explains the DGL spectrum.
Our result implies that the size distribution is composed of small grains with a half-mass grain radius $a_{0.5} \approx 0.06 \mu$m.

\acknowledgments

This work was supported by NASA APRA research grants NNX07AI54G, NNG05WC18G, NNX07AG43G, NNX07AJ24G, and NNX10AE12G.
Initial support was provided by an award to J.B. from the Jet Propulsion Laboratory's Director's Research and Development Fund. 
CIBER was supported by KAKENHI (20·34, 18204018, 19540250, 21340047 and 21111004) from the Japan Society for the Promotion of Science (JSPS) and the Ministry of Education, Culture, Sports, Science and Technology (MEXT). 
Korean participation in CIBER was supported by the Pioneer Project from Korea Astronomy and Space science Institute (KASI).

We would like to acknowledge the dedicated efforts of the sounding rocket staff at the NASA Wallops Flight Facility and the White Sands Missile Range.
P.K. and M.Z. acknowledge support from a NASA Postdoctoral Fellowship, A.C. acknowledges support from an NSF CAREER award, and T.A. acknowledges support from the JSPS Research Fellowship for Young Scientists.
A.C. acknowledges support from an NSF CAREER award AST-0645427 and NSF
AST-1313319.
H.M.L acknowledges support from grant 2012R1A4A1028713.

We thank T.D. Brandt for kindly providing data and models.

\bibliographystyle{apj}
\bibliography{apj-jour,reference}

\end{document}